%% file: 1.main.tex
\documentclass[11pt]{article}

\usepackage[margin=1.0in]{geometry}

\usepackage{graphicx}
\usepackage{dsfont}
\usepackage{amsmath,amsthm}
\usepackage{amssymb}
\usepackage{amsfonts}       % blackboard math symbols
\usepackage{mathtools}
\usepackage{enumitem}
\usepackage{url}            % simple URL typesetting
\usepackage{booktabs}       % professional-quality tables
\usepackage{longtable}      % tables that break across pages
\usepackage{nicefrac}       % compact symbols for 1/2, etc.
\usepackage{microtype}      % microtypography
\usepackage{natbib}
\usepackage{doi}
\usepackage{tikz}
\usepackage{tikz-cd}
\usepackage{appendix}
\usepackage{comment}
\usetikzlibrary{arrows.meta,positioning}
\usepackage{titletoc}
\usepackage{xcolor}
\usepackage{hyperref}
\hypersetup{
    colorlinks=true,
    linkcolor=blue,
    citecolor=blue,
    urlcolor=blue,
    filecolor=blue
}
\definecolor{covernavy}  {RGB}{  0,  32,  96}
\definecolor{inputbg} {RGB}{248, 249, 252}
\usepackage[capitalize,nameinlink]{cleveref}
\usepackage{pdflscape}
\usepackage{array}
\usepackage{multirow}
\usepackage{ragged2e}

\theoremstyle{plain}

\theoremstyle{definition}

\theoremstyle{remark}

\crefname{assumption}{Assumption}{Assumptions}
\Crefname{assumption}{Assumption}{Assumptions}
\crefname{condition}{Condition}{Conditions}
\Crefname{condition}{Condition}{Conditions}
\crefname{remark}{Remark}{Remarks}
\Crefname{remark}{Remark}{Remarks}

\begin{document}
\title{Causal Mediation Analysis for an Interrupted Time Series:\\
Stabilized Mediator Weighting with an Application to a Vehicle\\
Emissions Policy}

\author{
\textbf{Shalini Jayanetti}\\
Department of Statistics,\\
University of Manitoba\\
\href{mailto:jayanesk@myumanitoba.ca}{jayanesk@myumanitoba.ca}
\and
\textbf{Sumeet Kalia}\\
Department of Statistics,\\
University of Manitoba\\
\href{mailto:Sumeet.Kalia@umanitoba.ca}{Sumeet.Kalia@umanitoba.ca}
}

\renewcommand{\thefootnote}{\ensuremath{\dagger}}
\maketitle
\renewcommand{\thefootnote}{\arabic{footnote}}
\setcounter{footnote}{0}

\begin{abstract}
Population-level policies are introduced at a fixed time and evaluated from a
single series of aggregate outcomes, and the interrupted time series design
estimates the total shift in an outcome after the intervention. When the policy
is expected to act through a measurable pathway, the total effect is less
informative than its decomposition into direct and indirect effects. We formulate
causal mediation for a single interrupted time series and study stabilized
mediator weighting as the estimator of the natural direct and indirect effects.
Because the intervention is a deterministic function of calendar time, the
exposure weight equals one and the exposure contrast is identified through the
segmented-regression level shift, so only the mediator pathway is weighted. We add
a cumulative mediator weight that carries the lagged confounder history,
incorporate a concurrent event as a second interruption, and replace variance
formulas that treat the estimated weights as fixed with a block-residual bootstrap
that keeps the deterministic exposure timing intact and resamples the mediator
and outcome residuals in moving blocks. In a simulation calibrated to daily data,
the unweighted product-of-coefficients estimator is biased under
mediator-outcome confounding, with an indirect-effect bias near $0.19$ and
coverage of $0.003$, whereas stabilized weighting reduces the bias to about $0.03$
and improves indirect-effect coverage from near zero to about $0.83$. Applied to
the 2019 termination of Ontario's Drive Clean vehicle emissions testing program,
the method estimates a direct reduction in ground-level ozone of $2.113$ parts per
billion (95\% interval $-3.384$ to $-0.841$) that is robust across four Toronto
regions, a positive but heterogeneous indirect effect through nitrogen dioxide,
and a total effect near the boundary of significance; a pre-pandemic sensitivity
analysis agrees with the primary results. The estimated direction of the direct
effect is consistent with associations reported in the atmospheric-science
literature.
 
\end{abstract}

%\newpage
% \tableofcontents

\newpage

\input{2.body}

\bibliographystyle{apalike}
\bibliography{references}

\newpage
\begin{appendices}
\input{3.appendix}
\end{appendices}

\end{document}

%% file: 2.body.tex
\section{Introduction}
\label{sec:intro}

Population-level interventions are often introduced at a known date and evaluated
from a single series of aggregate outcomes. The interrupted time series (ITS)
design with segmented regression is the standard framework for such evaluations,
because it estimates the level and slope change in an outcome after an
intervention while adjusting for the pre-intervention trend, seasonality, and
concurrent events \citep{wagner2002segmented, bernal2017its, bhaskaran2013time}.
Segmented regression summarizes the total change in the outcome, but it does not
distinguish the part of that change that is transmitted through an intermediate
variable from the part that is not. When an intervention is expected to operate
through a specific pathway, the total change is a coarse summary, and a
decomposition into direct and indirect components is more informative for
mechanism and for policy.

Causal mediation analysis provides that decomposition through the natural direct
effect (NDE) and the natural indirect effect (NIE), defined with nested
counterfactuals in the potential-outcomes framework
\citep{robins1992identifiability, pearl2001direct}. The product-of-coefficients
procedure of \citet{baron1986moderator} remains in wide use, but it estimates
valid causal quantities only under conditions that were not part of its original
justification. Modern treatments make these conditions explicit and separate the
controlled direct effect from the natural direct and indirect effects
\citep{imai2010general, vanderweele2015explanation, vanderweele2016practitioner}.
The identifying assumptions are consistency, exchangeability for the exposure and
for the mediator, positivity, and the absence of a mediator-outcome confounder
that is itself affected by the exposure; under these assumptions the natural
effects are given by the mediation formula \citep{pearl2001direct, imai2010general}.

Regression-based estimators specify models for the mediator and the outcome and combine them
through the mediation formula \citep{valeri2013mediation}. Simulation and
g-computation estimators integrate over the mediator distribution
\citep{imai2010general}. Weighting estimators construct a pseudo-population in
which the mediator is independent of the measured confounders and then fit a
marginal structural model for the natural effects
\citep{robins2000marginal, vanderweele2009msm, hong2010ratio, lange2012simple}.
The weighting approach is attractive when the mediator-outcome confounders are
numerous or high-dimensional, because it removes them through the weights rather
than through a correctly specified outcome model, and because the resulting effect
parameters are reported directly. Its central requirement is a correctly specified
model for the conditional density of the mediator given the exposure and the
confounders.

When confounders of the mediator-outcome relationship are themselves affected by prior exposure or
mediator values, the natural direct and indirect effects are not identified, and
only randomized interventional analogues can be estimated
\citep{vanderweele2014decomposition, vanderweele2017timevarying}. Estimators for
these settings include the parametric mediational g-formula
\citep{lin2017gformula}, weighting through paired marginal structural models
\citep{vanderweele2017timevarying}, and targeted minimum-loss estimation
\citep{zheng2012survival}. These methods presume repeated stochastic exposures
observed across many units, and they draw inference from independent replicates.

Another complication arises when the data are a single long series for one unit
rather than a sample of independent subjects. Causal estimands must then be
defined on the potential-outcome path of one unit, and identification and
inference rest on stationarity and weak temporal dependence in place of
independent sampling \citep{bojinov2019time, blackwell2018tscs}. The serial
correlation of daily observations has to be propagated into the standard errors,
which rules out variance formulas built for independent data.

Mediation estimators for time-varying
processes are designed for repeated stochastic exposures across many units and for
interventional analogues of the natural effects, and inference is drawn from
independent replicates. The ITS setting is different in three respects that change
both identification and inference. The exposure is a single deterministic switch
in calendar time, so its propensity is degenerate, positivity fails for the
exposure by construction, and the exposure weight collapses to one; the exposure
contrast is then identified through the segmented-regression level shift rather
than through weighting. Only the mediator, which is stochastic and confounded,
calls for weighting, so the estimator reduces to mediator weighting alone.
Inference must be drawn from one autocorrelated series, which requires resampling
that preserves both the deterministic exposure timing and the serial dependence.
Concurrent events during the post-intervention period, such as a later policy
change, act as additional interruptions that must be separated from the exposure
effect. The behavior of mediator weighting under this structure, and in particular
its sensitivity to the temporal dependence of the mediator's confounders and to
weight truncation, has not been characterized.

We formulate causal mediation for a single interrupted time series and study
stabilized mediator weighting as the estimator of the natural direct and indirect
effects. We state the estimand and the
identification conditions for the natural effects when the exposure is a
deterministic interruption, and we show that the exposure weight equals one and
that the exposure contrast is the segmented-regression level shift, so that
mediator weighting is the only weighting required
\citep{vanderweele2009msm, lange2012simple}. We adapt stabilized mediator
weighting to this setting and introduce a cumulative mediator weight that balances
the recent confounder path when the mediator's confounders are serially dependent.
We incorporate a concurrent post-intervention event as a second interruption, so
that the exposure effect is separated from the shift induced by that event. We
replace variance formulas that treat the estimated weights as fixed, and which
overstate coverage, with a block-residual bootstrap that holds the deterministic
exposure timing fixed and resamples the mediator and outcome residuals in moving
blocks, thereby propagating weight-estimation uncertainty and the serial
dependence of daily data \citep{kunsch1989jackknife, lahiri2003resampling}. We
evaluate the estimators in a simulation study calibrated to daily data, with
autocorrelated continuous and binary confounders, g-computation truth, Monte Carlo
standard errors on every performance measure \citep{morris2019using}, and
bootstrap coverage. We apply the method to the 2019 termination of Ontario's Drive
Clean vehicle emissions testing program, estimating its effect on ground-level
ozone through nitrogen dioxide across four Toronto regions, and we report
region-specific estimates with a random-effects pooled summary.

\section{Notation and causal framework}
\label{sec:framework}

Let $t = 1, \dots, n$ index consecutive days for one geographic unit. The observed
data on day $t$ are $\mathcal{O}_t = (X_t, A_t, M_t, Y_t)$, where $Y_t$ is the
outcome, $M_t$ is the mediator, $A_t \in \{0, 1\}$ is the exposure indicator, and
$X_t$ collects the measured time-varying confounders together with their recent
lags. The exposure is a deterministic function of calendar time,
$A_t = \mathbf{1}\{t \ge t_0\}$, where $t_0$ is the day the intervention takes
effect. A second known event at day $t_1 > t_0$ is recorded by
$C_t = \mathbf{1}\{t \ge t_1\}$ and enters as a concurrent interruption.

Let $M_t(a)$ denote the mediator that would be observed under exposure level $a$,
and let $Y_t(a, m)$ denote the outcome that would be observed under exposure $a$
and mediator value $m$ \citep{robins1992identifiability, pearl2001direct}. The
natural direct and indirect effects, evaluated at a representative
post-intervention day and taken to be constant over the post-intervention period,
are
\begin{align}
\mathrm{NDE} &= \mathbb{E}\!\left[\, Y_t\big(1, M_t(0)\big) - Y_t\big(0, M_t(0)\big)\,\right], \label{eq:nde}\\
\mathrm{NIE} &= \mathbb{E}\!\left[\, Y_t\big(1, M_t(1)\big) - Y_t\big(1, M_t(0)\big)\,\right], \label{eq:nie}
\end{align}
and the total effect is $\mathrm{TE} = \mathrm{NDE} + \mathrm{NIE}$. Constancy of
the effects over the post-period is a working assumption; a time-varying
formulation is noted in Section~\ref{sec:disc}.

Identification of \eqref{eq:nde} and \eqref{eq:nie} requires the following
conditions, stated at the population level for the single-unit process.
Consistency links the observed and potential quantities, so that $M_t = M_t(A_t)$
and $Y_t = Y_t(A_t, M_t)$. Because the exposure is deterministic in calendar time,
its propensity is degenerate, exposure positivity fails, and the exposure weight
is fixed at one; the exposure contrast is identified through the
segmented-regression level shift rather than through weighting. The mediator
pathway requires exchangeability for the mediator, $Y_t(a, m) \perp M_t \mid A_t,
X_t$, so that the confounders in $X_t$ account for the association between the
mediator and the outcome that does not reflect the mediator's effect.
Identification of the natural effects also requires that no confounder of the
mediator-outcome relationship is itself affected by the exposure
\citep{tchetgen2014natural, vanderweele2014decomposition}; when such a confounder
is present, the natural effects are not identified and only interventional
analogues are \citep{vanderweele2017timevarying}. In the application the
confounders in $X_t$ are meteorological variables that are not affected by the
local intervention, so this condition is plausible. The mediator further requires
positivity, so that the conditional density of $M_t$ given $(A_t, X_t)$ has common
support over the region used in weighting \citep{ding2024first}. Inference from a
single series requires stationarity and weak temporal dependence in place of
independent sampling \citep{bojinov2019time, blackwell2018tscs}.

In the pseudo-population induced by the mediator weight, the mediator is
independent of $X_t$ given the exposure, and a regression of $Y_t$ on $A_t$ and
$M_t$ recovers the direct and indirect effects. The coefficients of that
regression are population-averaged causal parameters in the weighted
pseudo-population and are not conditional regression coefficients.

\section{Methods}
\label{sec:methods}

\subsection{Segmented-regression model with a concurrent interruption}
\label{sec:seg}

The outcome follows a segmented regression in which the intervention and the
concurrent event each contribute a level change \citep{bernal2017its}. Writing
$S_t$ for smooth seasonal terms and $D_t$ for a day-type indicator, the outcome
model in the weighted pseudo-population is
\begin{equation}
\mathbb{E}_w[\, Y_t \mid A_t, M_t, C_t \,]
  = \delta_0 + \delta_1 A_t + \delta_2 M_t + \delta_c C_t + \delta_d D_t + S_t.
\label{eq:outcome}
\end{equation}
The exposure-induced level shift in the mediator is estimated from
\begin{equation}
\mathbb{E}[\, M_t \mid A_t, C_t \,] = \alpha_0 + \alpha_1 A_t + \alpha_c C_t,
\label{eq:mediator}
\end{equation}
which is unweighted because the exposure is deterministic. Under the linear
working models the natural effects are $\mathrm{NDE} = \delta_1$ and
$\mathrm{NIE} = \alpha_1 \delta_2$, with $\mathrm{TE} = \delta_1 + \alpha_1
\delta_2$. This is the product-of-coefficients form of \citet{baron1986moderator}
with the outcome regression weighted to remove mediator-outcome confounding. The
day type $D_t$ is a predictor of the outcome and is not a confounder of the
mediator-outcome relationship, so it enters the outcome model \eqref{eq:outcome}
for precision and is excluded from the weights.

\subsection{Mediator weighting}
\label{sec:weights}

The exposure weight is fixed at one. The stabilized mediator weight on day $t$ is
the ratio of two conditional densities of the mediator,
\begin{equation}
w^M_t = \frac{f\!\left(M_t \mid A_t, C_t\right)}
             {f\!\left(M_t \mid A_t, C_t, X_t\right)},
\label{eq:medweight}
\end{equation}
where both densities are Gaussian with means from linear models and a common-form
variance. The numerator conditions on the variables retained in the outcome model,
which stabilizes the weight and controls its variance, and the denominator adds
the confounder history in $X_t$. Weighting the outcome regression by $w^M_t$
yields a pseudo-population in which $M_t$ is independent of $X_t$ given $(A_t,
C_t)$, so the coefficient $\delta_2$ estimates the mediator-outcome effect free of
confounding by $X_t$ \citep{vanderweele2009msm, lange2012simple}.

When the mediator's confounders are serially dependent, the single-day weight does
not carry the recent confounder history. We therefore also use a cumulative
mediator weight over a short window,
\begin{equation}
w^{M,\mathrm{cum}}_t = w^M_t \, w^M_{t-1} \, w^M_{t-2},
\label{eq:cumweight}
\end{equation}
which balances the recent confounder path at the cost of a heavier-tailed weight
and a smaller effective sample size. Both weights are truncated at their 99th
percentile to limit the influence of extreme values \citep{robins2000marginal}.
The single-day weight \eqref{eq:medweight} is the primary specification, and the
cumulative weight \eqref{eq:cumweight} is a robustness specification.

\subsection{Inference}
\label{sec:inference}

Variance formulas that treat the estimated weights as fixed understate the
variability of the weighted estimator and overstate its coverage, and the delta
method for the product $\alpha_1 \delta_2$ ignores the joint estimation of the two
models. We therefore use a block-residual bootstrap that respects the
single-series structure. The design, comprising the exposure and concurrent-event
indicators, the confounders, and the day index, is held fixed, which preserves the
deterministic exposure timing. Rich generating models for the mediator and the
outcome are fitted once; their residuals are resampled in moving blocks 
mediator and outcome series are formed from the fitted values and the resampled
residual blocks; and the full estimation pipeline, comprising the weights, both
regressions, and the product term, is recomputed on each resample. Moving blocks
preserve the serial dependence of daily data, and refitting the weights on each
resample propagates weight-estimation uncertainty into the direct, indirect, and
total effects.

\subsection{Pooling across regions}
\label{sec:pool}

Region-specific estimates are the primary results. A random-effects summary across
regions is reported for description, with the between-region variance estimated by
restricted maximum likelihood and the bootstrap variances used as the
within-region variances \citep{dersimonian1986meta, veroniki2016methods}.
Heterogeneity is summarized by $I^2$, the proportion of total variation across
regions attributable to between-region variance rather than to within-region
sampling error \citep{higgins2003measuring}. With four regions the pooled estimate
is descriptive, and no moderator meta-regression is fitted.

\section{Simulation study}
\label{sec:sim}

\subsection{Design}
\label{sec:simdesign}

The design and reporting follow the guidance of \citet{morris2019using}. Each data
set is a single series of $n = 1200$ days. A continuous confounder and a binary
confounder each follow an AR(1) process with autoregressive parameter $\phi = 0.6$
and enter the mediator and the outcome at lags $0$, $1$, and $2$, so that the
mediator-outcome confounding is serially dependent. The exposure switches on at
$t_0 = 0.6n$. The mediator is
$M_t = 1 + \beta_1 A_t + \lambda \sum_{k=0}^{2}(X_{t-k} + B_{t-k}) + \varepsilon^M_t$,
and the outcome is
$Y_t = 1 + \gamma_1 A_t + \gamma_2 M_t + \lambda \sum_{k=0}^{2}(X_{t-k} + B_{t-k}) + \varepsilon^Y_t$,
with confounder coefficient $\lambda = 0.15$ and independent standard Gaussian
errors. Under the linear model the true effects are $\mathrm{NDE} = \gamma_1$ and
$\mathrm{NIE} = \beta_1 \gamma_2$, which we confirmed by g-computation on a series
of $2 \times 10^5$ days.

Three estimators are compared: the unweighted product-of-coefficients estimator
(Standard), the single-day stabilized mediator weight (sIPW), and the cumulative
mediator weight (Cum-IPW). Standard errors come from the block-residual bootstrap
with $B = 199$ draws, and the study uses $1000$ replicates. Performance is
summarized by bias, the empirical standard deviation (ESD) across replicates, the
mean squared error (MSE), and the coverage of nominal 95\% Wald intervals, each
reported with its Monte Carlo standard error.

The study comprises two sets of scenarios that share the same data-generating
process. The first set varies the effect structure, through the presence and
combination of the direct and indirect effects, while the nuisance hyperparameters
are held fixed. The second set holds the combined effect configuration
$(\beta_1, \gamma_1, \gamma_2) = (1, 0.25, 0.25)$ and varies one hyperparameter at
a time: the confounder coefficient $\lambda$, which sets the strength of the
mediator-outcome confounding and the resulting weight variability; the
autoregressive parameter $\phi$, which sets the serial dependence of the
confounders; the series length $n$; and the truncation quantile $q$, which governs
the bias-variance trade-off of weight truncation. The parameter configuration values are show in Table~\ref{tab:simdesign}.

\begin{table}[htbp]
\centering
\caption{Simulation design. The upper panel gives the effect configurations; the
lower panel gives the hyperparameter levels, varied one at a time about the
combined configuration $(\beta_1, \gamma_1, \gamma_2) = (1, 0.25, 0.25)$.}
\label{tab:simdesign}
\begin{tabular}{lccccc}
\toprule
\multicolumn{6}{l}{\textit{Effect configurations}} \\
\midrule
Scenario & $\beta_1$ & $\gamma_1$ & $\gamma_2$ & NDE & NIE \\
\midrule
Null      & 0 & 0.00 & 0.00 & 0.00 & 0.00 \\
Direct    & 0 & 0.25 & 0.00 & 0.25 & 0.00 \\
Indirect  & 1 & 0.00 & 0.25 & 0.00 & 0.25 \\
Both      & 1 & 0.25 & 0.25 & 0.25 & 0.25 \\
\bottomrule
\end{tabular}

\vspace{1em}

\begin{tabular}{lll}
\toprule
\multicolumn{3}{l}{\textit{Hyperparameter levels}} \\
\midrule
Hyperparameter & Base & Levels \\
\midrule
Confounder coefficient $\lambda$ & 0.15 & $0.05,\ 0.10,\ 0.20,\ 0.30$ \\
Autoregressive parameter $\phi$  & 0.60 & $0.00,\ 0.30,\ 0.60,\ 0.90$ \\
Series length $n$                & 1200 & $600,\ 1200,\ 2400,\ 4800$ \\
Truncation quantile $q$          & 0.99 & $0.95,\ 0.99,\ 1.00$ \\
\bottomrule
\end{tabular}
\end{table}

\subsection{Results}
\label{sec:simbase}

Table~\ref{tab:simresults} reports the results for the four effect configurations
over $1000$ replicates. In the null and direct scenarios all three estimators are
close to
unbiased, since there is no mediator-outcome confounding to remove. The
block-residual bootstrap intervals for the direct effect undercover in these
scenarios, with coverage between $0.83$ and $0.88$, and the intervals for the
near-zero indirect effect are conservative, with coverage above $0.98$. In the
indirect and combined scenarios the unweighted estimator is biased, with a
direct-effect bias near $-0.20$ and an indirect-effect bias near $+0.19$, and its
indirect-effect coverage falls to $0.003$. Stabilized mediator weighting reduces
both biases to about $0.03$ in absolute value and raises indirect-effect coverage
to about $0.83$; the cumulative weight gives similar bias with slightly higher
coverage, near $0.86$, at the cost of a larger empirical standard deviation.  The unweighted estimator is unusable for the indirect effect
under confounding, in both bias and coverage, and weighting corrects this. The
weighted intervals also undercover the direct effect, by an amount that is present
even without confounding, which indicates that the Wald block-bootstrap interval
is imperfect for these estimators; Section~\ref{sec:disc} returns to this point.
The Monte Carlo standard errors, at or below $0.004$ for bias and near $0.012$ for
coverage, confirm that the differences between estimators exceed simulation error.

\begin{table}[htbp]
\centering
\caption{Simulation results for the four effect configurations over $1000$
replicates. Bias, empirical standard deviation (ESD), mean squared error (MSE),
and coverage of nominal 95\% intervals, with Monte Carlo standard errors in
parentheses. True values in Table~\ref{tab:simdesign}.}
\label{tab:simresults}
\small
\begin{tabular}{llrrrrr}
\hline
Scenario & Estimator & Effect & Bias & ESD & MSE & Coverage \\
\hline
\multirow{6}{*}{Null}
 & Standard & NDE & $-0.0003$ (0.0026) & 0.0829 & 0.0069 & 0.828 (0.012) \\
 & Standard & NIE & $\phantom{-}0.0000$ (0.0006) & 0.0180 & 0.0003 & 0.811 (0.012) \\
 & sIPW     & NDE & $\phantom{-}0.0002$ (0.0030) & 0.0959 & 0.0092 & 0.828 (0.012) \\
 & sIPW     & NIE & $-0.0001$ (0.0001) & 0.0045 & 0.0000 & 0.989 (0.003) \\
 & Cum-IPW  & NDE & $\phantom{-}0.0002$ (0.0035) & 0.1101 & 0.0121 & 0.879 (0.010) \\
 & Cum-IPW  & NIE & $-0.0002$ (0.0002) & 0.0049 & 0.0000 & 0.991 (0.003) \\
\hline
\multirow{6}{*}{Direct}
 & Standard & NDE & $\phantom{-}0.0017$ (0.0026) & 0.0809 & 0.0065 & 0.852 (0.011) \\
 & Standard & NIE & $-0.0006$ (0.0006) & 0.0180 & 0.0003 & 0.801 (0.013) \\
 & sIPW     & NDE & $-0.0006$ (0.0030) & 0.0943 & 0.0089 & 0.835 (0.012) \\
 & sIPW     & NIE & $\phantom{-}0.0000$ (0.0001) & 0.0044 & 0.0000 & 0.987 (0.004) \\
 & Cum-IPW  & NDE & $-0.0020$ (0.0034) & 0.1076 & 0.0116 & 0.883 (0.010) \\
 & Cum-IPW  & NIE & $\phantom{-}0.0000$ (0.0002) & 0.0050 & 0.0000 & 0.985 (0.004) \\
\hline
\multirow{6}{*}{Indirect}
 & Standard & NDE & $-0.2000$ (0.0027) & 0.0850 & 0.0472 & 0.195 (0.013) \\
 & Standard & NIE & $\phantom{-}0.1977$ (0.0016) & 0.0520 & 0.0418 & 0.003 (0.002) \\
 & sIPW     & NDE & $-0.0342$ (0.0031) & 0.0987 & 0.0109 & 0.835 (0.012) \\
 & sIPW     & NIE & $\phantom{-}0.0339$ (0.0014) & 0.0440 & 0.0031 & 0.827 (0.012) \\
 & Cum-IPW  & NDE & $-0.0383$ (0.0037) & 0.1165 & 0.0150 & 0.867 (0.011) \\
 & Cum-IPW  & NIE & $\phantom{-}0.0357$ (0.0015) & 0.0486 & 0.0036 & 0.850 (0.011) \\
\hline
\multirow{6}{*}{Both}
 & Standard & NDE & $-0.1941$ (0.0027) & 0.0864 & 0.0451 & 0.222 (0.013) \\
 & Standard & NIE & $\phantom{-}0.1944$ (0.0016) & 0.0504 & 0.0403 & 0.003 (0.002) \\
 & sIPW     & NDE & $-0.0304$ (0.0032) & 0.1005 & 0.0110 & 0.835 (0.012) \\
 & sIPW     & NIE & $\phantom{-}0.0309$ (0.0014) & 0.0430 & 0.0028 & 0.853 (0.011) \\
 & Cum-IPW  & NDE & $-0.0328$ (0.0037) & 0.1185 & 0.0151 & 0.861 (0.011) \\
 & Cum-IPW  & NIE & $\phantom{-}0.0325$ (0.0015) & 0.0486 & 0.0034 & 0.862 (0.011) \\
\hline
\end{tabular}
\end{table}

Table~\ref{tab:simadd} reports the hyperparameter variations, with each cell giving
the bias and the empirical coverage. Varying the confounder coefficient $\lambda$
leaves the unweighted bias near $-0.19$ for the direct effect and $+0.19$ for the
indirect effect across the range, with coverage of the indirect effect at or below
$0.004$; the weighted estimators hold the bias near $0.03$ and coverage between
$0.81$ and $0.89$, so their performance is stable in the strength of the
confounding. Varying the autoregressive parameter $\phi$ has a large effect. At
$\phi = 0$ the unweighted bias is $-0.08$ with indirect-effect coverage $0.38$,
and at $\phi = 0.9$ the bias reaches $-0.50$ with coverage near zero; the weighted
estimators track the same pattern but far less severely, with coverage above $0.90$
up to $\phi = 0.6$ and a marked loss only at $\phi = 0.9$, where the cumulative
weight is less biased than the single-day weight ($-0.154$ against $-0.179$ for the
direct effect) and better covered ($0.681$ against $0.496$). Varying the series
length $n$ exposes the two estimators' different limits: the unweighted
direct-effect coverage falls from $0.44$ at $n = 600$ to $0.003$ at $n = 4800$, the
signature of a bias that does not shrink, while the weighted estimators keep a bias
near $0.03$ but see their coverage decline with $n$ as the residual truncation bias
becomes large relative to the shrinking standard error. Varying the truncation
quantile $q$ isolates that residual bias: at $q = 1.00$, with no truncation, the
single-day bias falls to $-0.006$ for the direct effect and $0.003$ for the
indirect effect, with coverage $0.87$ and $0.93$, whereas heavier truncation at
$q = 0.95$ raises the bias and lowers the coverage. The residual bias of the
weighted estimators is therefore largely a truncation artifact, which motivates the
doubly robust estimator discussed in Section~\ref{sec:disc}.

\begin{table}[htbp]
\centering
\caption{Simulation results for the hyperparameter variations over $1000$
replicates, at the combined effect configuration
$(\beta_1, \gamma_1, \gamma_2) = (1, 0.25, 0.25)$. Each cell reports the bias with
the empirical coverage of nominal 95\% intervals in parentheses. Each panel varies
one hyperparameter at the levels shown; the base values are $\lambda = 0.15$,
$\phi = 0.6$, $n = 1200$, and $q = 0.99$.}
\label{tab:simadd}
\footnotesize
\begin{tabular}{ll cccc}
\toprule
Estimator & Effect & \multicolumn{4}{c}{Level} \\
\midrule
\multicolumn{6}{l}{\textit{Confounder coefficient} $\lambda$: $0.05,\ 0.10,\ 0.20,\ 0.30$} \\
Standard & NDE & $-0.189$ (0.240) & $-0.192$ (0.237) & $-0.192$ (0.245) & $-0.197$ (0.211) \\
Standard & NIE & $\phantom{-}0.193$ (0.004) & $\phantom{-}0.194$ (0.003) & $\phantom{-}0.193$ (0.000) & $\phantom{-}0.193$ (0.003) \\
sIPW     & NDE & $-0.025$ (0.835) & $-0.027$ (0.835) & $-0.031$ (0.810) & $-0.035$ (0.820) \\
sIPW     & NIE & $\phantom{-}0.029$ (0.854) & $\phantom{-}0.030$ (0.854) & $\phantom{-}0.031$ (0.834) & $\phantom{-}0.031$ (0.862) \\
Cum-IPW  & NDE & $-0.027$ (0.888) & $-0.025$ (0.868) & $-0.033$ (0.858) & $-0.033$ (0.876) \\
Cum-IPW  & NIE & $\phantom{-}0.032$ (0.865) & $\phantom{-}0.030$ (0.857) & $\phantom{-}0.033$ (0.859) & $\phantom{-}0.032$ (0.874) \\
\midrule
\multicolumn{6}{l}{\textit{Autoregressive parameter} $\phi$: $0.00,\ 0.30,\ 0.60,\ 0.90$} \\
Standard & NDE & $-0.081$ (0.744) & $-0.115$ (0.573) & $-0.199$ (0.219) & $-0.496$ (0.013) \\
Standard & NIE & $\phantom{-}0.079$ (0.376) & $\phantom{-}0.114$ (0.101) & $\phantom{-}0.192$ (0.001) & $\phantom{-}0.485$ (0.018) \\
sIPW     & NDE & $-0.010$ (0.926) & $-0.014$ (0.907) & $-0.037$ (0.815) & $-0.179$ (0.496) \\
sIPW     & NIE & $\phantom{-}0.009$ (0.944) & $\phantom{-}0.013$ (0.944) & $\phantom{-}0.030$ (0.874) & $\phantom{-}0.168$ (0.243) \\
Cum-IPW  & NDE & $-0.009$ (0.937) & $-0.015$ (0.922) & $-0.036$ (0.882) & $-0.154$ (0.681) \\
Cum-IPW  & NIE & $\phantom{-}0.009$ (0.944) & $\phantom{-}0.014$ (0.940) & $\phantom{-}0.031$ (0.876) & $\phantom{-}0.144$ (0.407) \\
\midrule
\multicolumn{6}{l}{\textit{Series length} $n$: $600,\ 1200,\ 2400,\ 4800$} \\
Standard & NDE & $-0.194$ (0.439) & $-0.196$ (0.198) & $-0.194$ (0.043) & $-0.196$ (0.003) \\
Standard & NIE & $\phantom{-}0.194$ (0.060) & $\phantom{-}0.196$ (0.003) & $\phantom{-}0.196$ (0.000) & $\phantom{-}0.196$ (0.000) \\
sIPW     & NDE & $-0.034$ (0.830) & $-0.033$ (0.847) & $-0.029$ (0.815) & $-0.031$ (0.779) \\
sIPW     & NIE & $\phantom{-}0.034$ (0.883) & $\phantom{-}0.033$ (0.838) & $\phantom{-}0.031$ (0.770) & $\phantom{-}0.031$ (0.614) \\
Cum-IPW  & NDE & $-0.036$ (0.884) & $-0.033$ (0.887) & $-0.027$ (0.873) & $-0.033$ (0.830) \\
Cum-IPW  & NIE & $\phantom{-}0.036$ (0.891) & $\phantom{-}0.034$ (0.859) & $\phantom{-}0.032$ (0.806) & $\phantom{-}0.032$ (0.668) \\
\midrule
\multicolumn{6}{l}{\textit{Truncation quantile} $q$: $0.95,\ 0.99,\ 1.00$} \\
Standard & NDE & $-0.198$ (0.212) & $-0.195$ (0.222) & $-0.198$ (0.228) & \\
Standard & NIE & $\phantom{-}0.194$ (0.005) & $\phantom{-}0.196$ (0.004) & $\phantom{-}0.193$ (0.004) & \\
sIPW     & NDE & $-0.065$ (0.752) & $-0.030$ (0.835) & $-0.006$ (0.874) & \\
sIPW     & NIE & $\phantom{-}0.061$ (0.589) & $\phantom{-}0.033$ (0.860) & $\phantom{-}0.003$ (0.930) & \\
Cum-IPW  & NDE & $-0.061$ (0.806) & $-0.031$ (0.872) & $-0.015$ (0.910) & \\
Cum-IPW  & NIE & $\phantom{-}0.055$ (0.691) & $\phantom{-}0.034$ (0.865) & $\phantom{-}0.008$ (0.930) & \\
\bottomrule
\end{tabular}
\end{table}

\section{Application: termination of Ontario's Drive Clean program}
\label{sec:app}

\subsection{Data and design}
\label{sec:appdata}

Ontario ended light-duty vehicle emissions testing under the Drive Clean program
on April 1, 2019 \citep{driveclean2019}. We take this date as the interruption,
$A_t = \mathbf{1}\{t \ge \text{2019-04-01}\}$, and estimate its effect on
ground-level ozone (O$_3$) operating through nitrogen dioxide (NO$_2$). Daily mean
NO$_2$ and O$_3$ were computed from hourly monitor readings for four Toronto
regions, denoted Downtown, East, West, and North, over 2015 to 2024. Daily mean
temperature and total precipitation from Toronto Pearson, each at lags $1$ and
$2$, form the confounder history $X_t$, and a weekday-and-holiday indicator is the
day type. The onset of the Ontario declaration of emergency on March 17, 2020
enters as a concurrent interruption, $C_t = \mathbf{1}\{t \ge \text{2020-03-17}\}$,
so that the intervention effect is separated from the change in traffic and
emissions during the pandemic period. Analyses use between $3458$ and $3561$ days
per region. A background description of the data, a study-specific causal diagram,
and exploratory summaries are given in Appendix~\ref{app:background}.

The direction of the estimated effects is interpreted with reference to the
atmospheric-science literature rather than derived from a chemical model. In dense
urban settings, ozone concentrations have been reported to respond in the opposite
direction to changes in nitrogen oxides, so that higher local nitrogen oxide is
associated with lower ozone and lower nitrogen oxide with higher ozone
\citep{sillman1999relation, kleinman2000ozone}. We treat this only as context for
the estimated signs and do not model the mechanism.

\subsection{Estimation}
\label{sec:appest}

For each region the direct, indirect, and total effects are estimated by the three
methods of Section~\ref{sec:methods}. Standard errors and percentile intervals come
from the block-residual bootstrap with $500$ draws and a block length of $16$ days.
Region-specific estimates are pooled by restricted-maximum-likelihood random
effects for description.

\subsection{Results}
\label{sec:appresults}

Table~\ref{tab:appresults} reports the region-specific estimates under the
single-day weight in its upper panel and the pooled estimates for the three
methods in its lower panel. Under stabilized weighting the direct effect on ozone
is negative in every region and pools to $-2.113$ parts per billion (95\% interval
$-3.384$ to $-0.841$), with no detectable between-region heterogeneity
($I^2 \approx 0$). The indirect effect through NO$_2$ is positive in three regions
and negative in the Downtown core, which produces high heterogeneity in the
indirect pathway ($I^2 = 92.9\%$) and a pooled indirect effect of $0.761$ ppb
($-0.114$ to $1.635$) whose interval includes zero. The total effect pools to
$-1.307$ ppb ($-2.625$ to $0.010$). The cumulative weight attenuates all three
effects, consistent with the larger loss of effective sample size reported in the
diagnostics of Appendix~\ref{app:diag}.

The pre-pandemic sensitivity analysis restricts the post-period to April 2019
through mid-March 2020 and drops the concurrent interruption. It gives a pooled
direct effect of $-1.894$ ppb and a total effect of $-1.180$ ppb under stabilized
weighting, close to the primary estimates, so the direct and total effects are not
artifacts of the pandemic period. The concurrent-event coefficients act in
opposite directions on the two pollutants, with a negative coefficient on NO$_2$
in three of the four regions and a positive coefficient on ozone
(Appendix~\ref{app:diag}); this pattern matches the associations reported for
urban ozone during the 2020 restrictions
\citep{sicard2020amplified, adams2020air}.

\begin{table}[htbp]
\centering
\caption{Application estimates (ppb) with 95\% block-bootstrap percentile
intervals. The upper panel gives region-specific estimates under single-day
stabilized mediator weighting. The lower panel gives pooled random-effects
estimates for the three methods, under the primary analysis, which adjusts for the
pandemic interruption over 2015 to 2024, and the pre-pandemic sensitivity
analysis.}
\label{tab:appresults}
\footnotesize
\setlength{\tabcolsep}{4pt}
\begin{tabular}{llccc}
\toprule
 & & NDE & NIE & TE \\
\midrule
\multicolumn{5}{l}{\textit{Region-specific, single-day weighting}} \\
Downtown & & $-2.292$ ($-5.489$, $-0.121$) & $-0.484$ ($-0.933$, $-0.131$) & $-2.776$ ($-5.983$, $-0.754$) \\
East     & & $-2.370$ ($-4.719$, $\phantom{-}0.043$) & $\phantom{-}1.535$ ($\phantom{-}1.047$, $\phantom{-}2.173$) & $-0.834$ ($-3.432$, $\phantom{-}1.798$) \\
West     & & $-1.039$ ($-3.693$, $\phantom{-}1.083$) & $\phantom{-}0.966$ ($\phantom{-}0.521$, $\phantom{-}1.504$) & $-0.073$ ($-2.749$, $\phantom{-}2.364$) \\
North    & & $-2.885$ ($-5.474$, $-0.083$) & $\phantom{-}1.089$ ($\phantom{-}0.523$, $\phantom{-}1.713$) & $-1.796$ ($-4.517$, $\phantom{-}0.990$) \\
\midrule
\multicolumn{5}{l}{\textit{Pooled, random effects}} \\
Primary      & Standard & $-2.227$ ($-3.412$, $-1.041$) & $\phantom{-}0.787$ ($-0.147$, $\phantom{-}1.720$) & $-1.397$ ($-2.694$, $-0.101$) \\
Primary      & sIPW     & $-2.113$ ($-3.384$, $-0.841$) & $\phantom{-}0.761$ ($-0.114$, $\phantom{-}1.635$) & $-1.307$ ($-2.625$, $\phantom{-}0.010$) \\
Primary      & Cum-IPW  & $-1.381$ ($-3.387$, $\phantom{-}0.626$) & $\phantom{-}0.662$ ($-0.118$, $\phantom{-}1.441$) & $-0.812$ ($-2.842$, $\phantom{-}1.217$) \\
Pre-pandemic & Standard & $-2.148$ ($-3.236$, $-1.060$) & $\phantom{-}0.711$ ($-0.284$, $\phantom{-}1.706$) & $-1.392$ ($-2.710$, $-0.074$) \\
Pre-pandemic & sIPW     & $-1.894$ ($-3.094$, $-0.694$) & $\phantom{-}0.680$ ($-0.265$, $\phantom{-}1.625$) & $-1.180$ ($-2.419$, $\phantom{-}0.060$) \\
Pre-pandemic & Cum-IPW  & $-1.271$ ($-3.234$, $\phantom{-}0.692$) & $\phantom{-}0.567$ ($-0.279$, $\phantom{-}1.414$) & $-0.766$ ($-2.760$, $\phantom{-}1.228$) \\
\bottomrule
\end{tabular}
\end{table}

\section{Discussion}
\label{sec:disc}

We formulated causal mediation for a single interrupted time series and studied
stabilized mediator weighting as the estimator of the natural direct and indirect
effects. When the exposure is a deterministic interruption the exposure weight
equals one and the exposure contrast is the segmented-regression level shift, so
mediator weighting is the only weighting the design admits. The simulation showed
that the unweighted product-of-coefficients estimator is biased and severely
undercovers the indirect effect whenever the mediator-outcome relationship is
confounded, and that mediator weighting removes most of that bias and restores much
of the lost coverage; the cumulative weight performs comparably at higher variance.
In the Drive Clean application, ending emissions testing was associated with a
reduction in ground-level ozone that operated directly rather than through nitrogen
dioxide, with a pooled direct effect near $-2.1$ ppb that was stable across the four
regions and to adjustment for the pandemic period.

The methodological contribution is to place the natural-effect decomposition within
the interrupted time series design and to characterize the estimator that this
design admits. Estimators of mediation effects for time-varying processes target
interventional analogues of the natural effects and draw inference from many
independent units \citep{vanderweele2017timevarying, lin2017gformula,
zheng2012survival}. The single-series interrupted time series collapses the exposure
weight to one, so mediator weighting is the sole weighting required
\citep{vanderweele2009msm, lange2012simple, hong2010ratio}, and it requires
inference from one autocorrelated series, which the block-residual bootstrap
supplies by resampling residual blocks while holding the deterministic timing fixed.
The cumulative mediator weight and the concurrent-event interruption address two
features of daily environmental series, the serial dependence of the mediator's
confounders and a later shock during the post-intervention period. The application
demonstrates that the aggregate effect of a policy on ozone can be misleading when
the direct and indirect pathways are not separated, since the indirect component
reverses sign in the urban core while the direct component does not.

The estimand and the identification
conditions are stated for the single-unit process, including the requirement that
no mediator-outcome confounder is affected by the exposure, a condition that is
often left implicit in applied mediation analyses \citep{tchetgen2014natural,
vanderweele2014decomposition}. Uncertainty is quantified by a resampling scheme
that preserves both the deterministic exposure timing and the serial dependence of
the data, rather than by variance formulas that treat the estimated weights as fixed
and overstate precision. The estimator evaluated in the simulation is the estimator
applied to the data, and every performance measure is reported with its Monte Carlo
standard error \citep{morris2019using}. The application treats the region-specific
estimates as primary, pools them only for description, and supports the main finding
with a pre-pandemic sensitivity analysis.

Stabilized mediator weighting is singly robust,
relying on a correctly specified model for the conditional density of the mediator,
and weight truncation trades a small bias for reduced variance; the simulation shows
that this residual bias is largely removed when truncation is relaxed, and that it
erodes coverage as the series lengthens and the standard error shrinks below the
bias. Serial dependence is the most demanding regime: at the strongest
autocorrelation examined, both the unweighted and the weighted estimators lose
accuracy, though the weighted estimators degrade far less and the cumulative weight
outperforms the single-day weight there. The block-bootstrap Wald intervals
undercover the direct effect even in the absence of confounding, which points to the
interval construction rather than the point estimator and indicates that a
studentized or longer-block bootstrap warrants study. The pooled application
estimate rests on four regions, so the between-region variance is estimated
imprecisely and reported descriptively, and the direction of the ozone effect is
interpreted through the atmospheric-science literature rather than a chemical model.

 A doubly robust
estimator for the single-series interrupted time series with a deterministic
exposure onset is the central open problem: multiply robust and
semiparametric-efficient estimators of the natural effects exist for point exposures
\citep{tchetgen2012semiparametric}, and g-formula, weighting, and targeted-learning
estimators of interventional analogues exist for time-varying exposures and
mediators \citep{lin2017gformula, vanderweele2017timevarying, zheng2012survival},
but none addresses the structure studied here. Allowing the direct and indirect
effects to vary over the post-intervention period, applying the graphical criteria
for mediation identification to a given series \citep{shpitser2011complete}, and
pooling more than four regions so that the between-region heterogeneity can be
estimated rather than described are natural extensions.

In summary, the natural-effect decomposition can be estimated for a single
interrupted time series when the exposure is deterministic, using mediator weighting
with a block-residual bootstrap. The method separates the direct and indirect
components of an intervention effect that a segmented regression reports only in
aggregate, and it recovers indirect effects that the unweighted
product-of-coefficients estimator does not.

\section{Acronyms}
\label{sec:acronyms}

{\footnotesize
\renewcommand{\arraystretch}{1}
\begin{longtable}{ll}
\toprule
\textbf{Acronym} & \textbf{Definition} \\
\midrule
\endfirsthead
\toprule
\textbf{Acronym} & \textbf{Definition} \\
\midrule
\endhead
\bottomrule
\endfoot
ADF     & Augmented Dickey-Fuller \\
AR(1)   & First-order autoregressive \\
CI      & Confidence interval \\
Cum-IPW & Cumulative inverse probability weighting \\
DAG     & Directed acyclic graph \\
DGP     & Data-generating process \\
ESD     & Empirical standard deviation \\
ESS     & Effective sample size \\
IPW     & Inverse probability weighting \\
ITS     & Interrupted time series \\
MC      & Monte Carlo \\
MSE     & Mean squared error \\
MSM     & Marginal structural model \\
NDE     & Natural direct effect \\
NIE     & Natural indirect effect \\
NO$_2$  & Nitrogen dioxide \\
NO$_x$  & Nitrogen oxides \\
O$_3$   & Ozone \\
REML    & Restricted maximum likelihood \\
sIPW    & Stabilized inverse probability weighting \\
SMD     & Standardized mean difference \\
TE      & Total effect \\
VOC     & Volatile organic compound \\
\end{longtable}
}

\section*{Acknowledgements}

This work was supported by Dr. Kalia's research grant and by funding from the
Department of Statistics and the Faculty of Science at the University of Manitoba.

\section*{Data Availability, Software, and Code}

The air-quality data are hourly NO$_2$ and O$_3$ monitor readings for four Toronto
regions, and the meteorological data are daily temperature and precipitation from
the Toronto Pearson station, all obtained from public environmental-monitoring
archives. The analysis uses base R \citep{rcore2024}; the random-effects pooling,
the density-ratio mediator weights, and the block-residual bootstrap are
implemented directly without additional packages. The R code that reproduces the
simulation study and the application, including the data-preparation steps, is
provided as a supplement, and the derived daily series can be reconstructed from
the code and the public archives.

%% file: 3.appendix.tex
\section{Identification of the natural effects under mediator weighting}
\label{app:ident}

This appendix restates the notation, gives the
identification conditions in full, and derives the mediator-weighting estimator of
the natural direct and indirect effects for a single interrupted time series.

\subsection*{Notation}

On day $t$ the observed data are $\mathcal{O}_t = (X_t, A_t, M_t, Y_t)$, where
$Y_t$ is the outcome, $M_t$ is the mediator, $A_t = \mathbf{1}\{t \ge t_0\}$ is the
deterministic exposure indicator, and $X_t$ collects the measured time-varying
confounders and their recent lags. A concurrent event at $t_1 > t_0$ is recorded
by $C_t = \mathbf{1}\{t \ge t_1\}$. The potential mediator under exposure $a$ is
$M_t(a)$, and the potential outcome under exposure $a$ and mediator $m$ is
$Y_t(a, m)$. The natural direct and indirect effects are given in \eqref{eq:nde}
and \eqref{eq:nie} and are taken to be constant over the post-intervention period.

\subsection*{Identification conditions}

The derivation uses the following conditions, stated at the population level for
the single-unit process. Consistency requires $M_t = M_t(A_t)$ and
$Y_t = Y_t(A_t, M_t)$. Because the exposure is a deterministic function of calendar
time, exposure positivity fails and the exposure weight is fixed at one; the
exposure contrast is identified through the segmented-regression level shift.
Exchangeability for the mediator requires $Y_t(a, m) \perp M_t \mid A_t, X_t$, so
that the confounders in $X_t$ account for the mediator-outcome association that is
not the mediator's effect. Identification of the natural effects also requires that
no confounder of the mediator-outcome relationship is affected by the exposure;
otherwise only interventional analogues are identified. Mediator positivity
requires that $f(M_t \mid A_t, C_t, X_t)$ has common support over the region used
in weighting. Inference relies on stationarity and weak temporal dependence.

\subsection*{Derivation}

Fix the exposure contrast at $a$ against $a'$ and hold the concurrent event at its
observed value. Under consistency and mediator exchangeability, the mean outcome
had the exposure been $a$ and the mediator been drawn from its distribution under
$a^\ast$ is
\begin{equation}
\mathbb{E}\!\left[ Y_t\big(a, M_t(a^\ast)\big) \right]
 = \int\!\!\int \mathbb{E}\!\left[ Y_t \mid A_t = a, M_t = m, X_t = x \right]\,
   f(m \mid a^\ast, x)\, f(x)\, dm\, dx,
\label{eq:mediationformula}
\end{equation}
which is the mediation formula \citep{pearl2001direct, imai2010general}. Setting
$a^\ast = a'$ gives the direct effect, and contrasting $a^\ast = a$ against $a'$
inside the mediator distribution gives the indirect effect.

The stabilized mediator weight
$w^M_t = f(M_t \mid A_t, C_t) / f(M_t \mid A_t, C_t, X_t)$ reweights the observed
joint density of $(M_t, X_t)$ given $(A_t, C_t)$,
\begin{equation}
f(m \mid a, c, x)\, f(x \mid a, c) \times w^M_t
 = f(m \mid a, c)\, f(x \mid a, c),
\label{eq:pseudopop}
\end{equation}
so that in the weighted pseudo-population the mediator is independent of the
confounders given the exposure and the concurrent event. The outer integral over
$x$ in \eqref{eq:mediationformula} then factors out, and the linear model
$\mathbb{E}_w[Y_t \mid A_t, M_t, C_t] = \delta_0 + \delta_1 A_t + \delta_2 M_t +
\delta_c C_t$ has coefficients equal to the direct effect $\delta_1$ and the
mediator-outcome effect $\delta_2$ free of confounding by $X_t$. Combining
$\delta_2$ with the mediator level shift $\alpha_1$ from
$\mathbb{E}[M_t \mid A_t, C_t] = \alpha_0 + \alpha_1 A_t + \alpha_c C_t$ gives
$\mathrm{NIE} = \alpha_1 \delta_2$ and $\mathrm{NDE} = \delta_1$ under the linear
working models, which is the product-of-coefficients estimator with the outcome
regression weighted by $w^M_t$. The exposure weight does not appear because the
exposure is deterministic. When the mediator's confounders are serially dependent,
the cumulative weight $w^M_t w^M_{t-1} w^M_{t-2}$ balances the recent confounder
path; both weights are truncated at their 99th percentile.

\section{Data-generating process for the simulation}
\label{app:dgp}

Table~\ref{tab:dgp} lists the parameters. Each data set is one series of $n$ days.
The continuous confounder $X_t$ and a latent variable $Z_t$ each follow an AR(1)
recursion with parameter $\phi$ and standard Gaussian innovations, and the binary
confounder is $B_t = \mathbf{1}\{Z_t > 0\}$. Both confounders enter the mediator
and the outcome at lags $0$, $1$, and $2$ with common coefficient $\lambda$. The
exposure switches on at $t_0 = 0.6n$. The mediator and outcome errors are
independent standard Gaussian. The true effects are $\mathrm{NDE} = \gamma_1$ and
$\mathrm{NIE} = \beta_1 \gamma_2$; a g-computation check on a series of
$2 \times 10^5$ days returned these values to within Monte Carlo error.

\begin{table}[htbp]
\centering
\caption{Simulation data-generating process. The last block lists the
hyperparameter levels used in the simulation study of
Section~\ref{sec:sim}.}
\label{tab:dgp}
\begin{tabular}{lll}
\hline
Symbol & Meaning & Value \\
\hline
$n$        & series length (days)                  & $1200$ \\
$t_0$      & exposure onset                        & $0.6n = 720$ \\
$\phi$     & AR(1) parameter of the confounders    & $0.6$ \\
$\lambda$  & confounder effect on $M_t$ and $Y_t$  & $0.15$ \\
$\beta_1$  & exposure effect on the mediator       & $\{0, 0, 1, 1\}$ \\
$\gamma_1$ & direct exposure effect on the outcome & $\{0, 0.25, 0, 0.25\}$ \\
$\gamma_2$ & mediator effect on the outcome        & $\{0, 0, 0.25, 0.25\}$ \\
$B$        & bootstrap draws per replicate         & $199$ \\
$n_{\text{MC}}$ & Monte Carlo replicates           & $1000$ \\
trim       & weight truncation quantile            & $0.99$ \\
\hline
$\lambda$ levels & additional scenarios            & $0.05, 0.10, 0.20, 0.30$ \\
$\phi$ levels    & additional scenarios            & $0.00, 0.30, 0.60, 0.90$ \\
$n$ levels       & additional scenarios            & $600, 1200, 2400, 4800$ \\
$q$ levels       & additional scenarios            & $0.95, 0.99, 1.00$ \\
\hline
\end{tabular}
\end{table}

\section{Multiple-time-point causal diagram}
\label{app:dag}

\begin{figure}[htbp]
\centering
\begin{tikzpicture}[>=Stealth, node distance=16mm,
  every node/.style={font=\small},
  var/.style={circle, draw, minimum size=8mm, inner sep=1pt},
  det/.style={circle, draw, dashed, minimum size=8mm, inner sep=1pt}]
  \node[var] (Xt) {$X_t$};
  \node[det, right=22mm of Xt] (At) {$A_t$};
  \node[var, below=12mm of Xt] (Mt) {$M_t$};
  \node[var, below=12mm of At] (Yt) {$Y_t$};
  \node[var, right=34mm of At] (Xt1) {$X_{t+1}$};
  \node[det, right=22mm of Xt1] (At1) {$A_{t+1}$};
  \node[var, below=12mm of Xt1] (Mt1) {$M_{t+1}$};
  \node[var, below=12mm of At1] (Yt1) {$Y_{t+1}$};
  \draw[->] (Xt) -- (Mt);
  \draw[->] (Xt) to[bend right=10] (Yt);
  \draw[->] (At) -- (Mt);
  \draw[->] (At) -- (Yt);
  \draw[->] (Mt) -- (Yt);
  \draw[->] (Xt) to[bend left=30] (Xt1);
  \draw[->] (Xt) to[bend left=-5] (Mt1);
  \draw[->] (Xt1) -- (Mt1);
  \draw[->] (Xt1) to[bend right=10] (Yt1);
  \draw[->] (At1) -- (Mt1);
  \draw[->] (At1) -- (Yt1);
  \draw[->] (Mt1) -- (Yt1);
\end{tikzpicture}
\caption{Causal diagram across two days. Dashed nodes mark the deterministic
exposure. The confounder $X_t$ confounds the mediator-outcome relationship within
a day and persists to the next day.}
\label{fig:dag}
\end{figure}
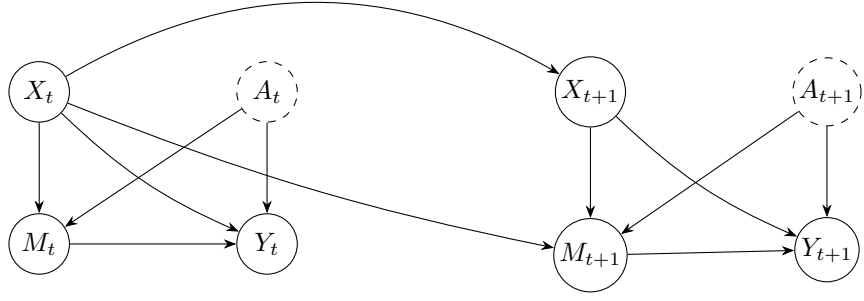

Figure~\ref{fig:dag} shows the structure across two consecutive days. The
confounder $X_t$ affects the mediator $M_t$ and the outcome $Y_t$ on the same day
and persists to the next day, which is why a single-day mediator weight can leave
residual imbalance and a cumulative weight over the recent window is used. The
exposure $A_t$ is deterministic in calendar time, indicated by the dashed node, so
no arrow enters it from the confounders. The mediator-outcome relationship is
confounded by $X_t$, and the mediator weight blocks the path
$M_t \leftarrow X_t \rightarrow Y_t$.

\section{Background on the Toronto ozone data}
\label{app:background}

\subsection{Conceptual framework}
\label{app:concept}

\begin{figure}[htbp]
\centering
\begin{tikzpicture}[>=Stealth, node distance=16mm,
  every node/.style={font=\small},
  var/.style={draw, rounded corners, minimum height=8mm, inner sep=3pt},
  det/.style={draw, dashed, rounded corners, minimum height=8mm, inner sep=3pt}]
  \node[var] (X) {Weather $X_t$};
  \node[det, right=26mm of X] (A) {Drive Clean removal $A_t$};
  \node[det, below=16mm of X] (C) {Pandemic onset $C_t$};
  \node[var, right=26mm of C] (M) {NO$_2$ $M_t$};
  \node[var, below=16mm of M] (Y) {Ozone $Y_t$};
  \draw[->] (X) to[bend left=8] (M);
  \draw[->] (X) to[bend right=2] (Y);
  \draw[->] (A) -- (M);
  \draw[->] (A) to[bend left=25] (Y);
  \draw[->] (C) -- (M);
  \draw[->] (C) to[bend right=18] (Y);
  \draw[->] (M) -- (Y);
\end{tikzpicture}
\caption{Study-specific causal diagram for the Drive Clean application. Dashed
nodes mark the deterministic interruptions, the removal of the Drive Clean program
and the onset of the pandemic restrictions. Nitrogen dioxide mediates the effect of
the policy on ozone, and the meteorological variables confound the
nitrogen-dioxide-to-ozone relationship.}
\label{fig:dag-toronto}
\end{figure}
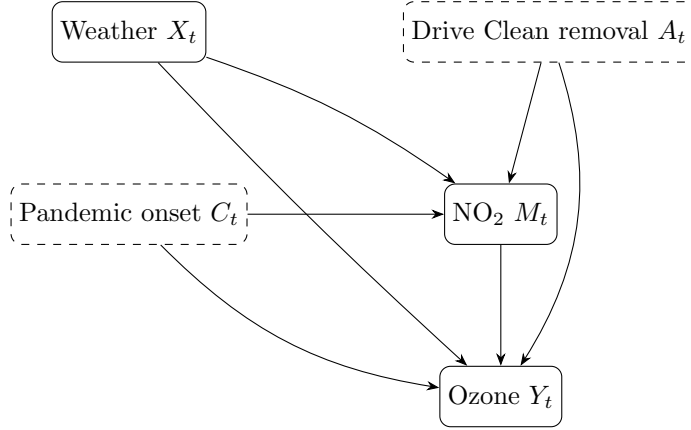

Figure~\ref{fig:dag-toronto} adapts the diagram of Figure~\ref{fig:dag} to the
application. The exposure is the removal of the Drive Clean program, denoted $A_t$,
which is deterministic in calendar time. The mediator is the daily nitrogen
dioxide level, denoted $M_t$, and the outcome is the daily ozone level, denoted
$Y_t$. The confounders $X_t$ are the meteorological variables, daily temperature
and precipitation with their recent lags, which affect both nitrogen dioxide and
ozone and persist across days. The onset of the pandemic restrictions, denoted
$C_t$, is a second deterministic interruption that affects both pollutants. The
meteorological confounders are not affected by the local policy, and neither
deterministic interruption receives an incoming arrow.

\subsection{Exploratory data analysis}
\label{app:eda}

Table~\ref{tab:eda-balance} reports the covariate balance between the pre-exposure
and post-exposure periods for each region. Because the exposure is deterministic
the exposure weights are one, so the comparison is between the two calendar periods
and the standardized mean difference (SMD) measures the imbalance across the
interruption. The meteorological confounders show small imbalances, with
temperature SMDs between $0.14$ and $0.17$ and precipitation SMDs below $0.06$,
which reflects the different calendar span of the post-period, and the day-type
indicator is balanced. The mediator differs across regions: nitrogen dioxide is
almost unchanged in the Downtown core (SMD $0.064$) but lower in the post-period in
the East, West, and North regions (SMD between $-0.490$ and $-0.388$), which
anticipates the region-specific indirect effects.

\begin{table}[htbp]
\centering
\caption{Covariate balance between the pre-exposure and post-exposure periods by
region. Entries are period means and the standardized mean difference (SMD).}
\label{tab:eda-balance}
\footnotesize
\begin{tabular}{lccc}
\toprule
Variable & Pre-exposure mean & Post-exposure mean & SMD \\
\midrule
\multicolumn{4}{l}{\textit{Downtown}} \\
Temperature          & 8.629 & 10.157 & $\phantom{-}0.144$ \\
Temperature, lag 1   & 8.613 & 10.155 & $\phantom{-}0.146$ \\
Temperature, lag 2   & 8.601 & 10.149 & $\phantom{-}0.146$ \\
Precipitation        & 2.065 & 2.392  & $\phantom{-}0.055$ \\
Precipitation, lag 1 & 2.064 & 2.388  & $\phantom{-}0.055$ \\
Precipitation, lag 2 & 2.074 & 2.433  & $\phantom{-}0.060$ \\
Weekday indicator    & 0.693 & 0.689  & $-0.009$ \\
NO$_2$ (mediator)    & 12.963 & 13.305 & $\phantom{-}0.064$ \\
\midrule
\multicolumn{4}{l}{\textit{East}} \\
Temperature          & 8.637 & 10.352 & $\phantom{-}0.162$ \\
Temperature, lag 1   & 8.625 & 10.345 & $\phantom{-}0.163$ \\
Temperature, lag 2   & 8.620 & 10.333 & $\phantom{-}0.162$ \\
Precipitation        & 2.061 & 2.386  & $\phantom{-}0.055$ \\
Precipitation, lag 1 & 2.065 & 2.382  & $\phantom{-}0.054$ \\
Precipitation, lag 2 & 2.075 & 2.396  & $\phantom{-}0.054$ \\
Weekday indicator    & 0.690 & 0.689  & $-0.001$ \\
NO$_2$ (mediator)    & 12.163 & 9.464 & $-0.490$ \\
\midrule
\multicolumn{4}{l}{\textit{West}} \\
Temperature          & 8.692 & 10.379 & $\phantom{-}0.160$ \\
Temperature, lag 1   & 8.678 & 10.374 & $\phantom{-}0.161$ \\
Temperature, lag 2   & 8.664 & 10.361 & $\phantom{-}0.161$ \\
Precipitation        & 2.067 & 2.393  & $\phantom{-}0.056$ \\
Precipitation, lag 1 & 2.066 & 2.387  & $\phantom{-}0.055$ \\
Precipitation, lag 2 & 2.077 & 2.410  & $\phantom{-}0.056$ \\
Weekday indicator    & 0.695 & 0.692  & $-0.007$ \\
NO$_2$ (mediator)    & 15.644 & 12.804 & $-0.479$ \\
\midrule
\multicolumn{4}{l}{\textit{North}} \\
Temperature          & 8.562 & 10.384 & $\phantom{-}0.172$ \\
Temperature, lag 1   & 8.545 & 10.381 & $\phantom{-}0.174$ \\
Temperature, lag 2   & 8.537 & 10.369 & $\phantom{-}0.174$ \\
Precipitation        & 2.147 & 2.370  & $\phantom{-}0.037$ \\
Precipitation, lag 1 & 2.125 & 2.383  & $\phantom{-}0.043$ \\
Precipitation, lag 2 & 2.135 & 2.398  & $\phantom{-}0.044$ \\
Weekday indicator    & 0.696 & 0.693  & $-0.007$ \\
NO$_2$ (mediator)    & 11.650 & 9.397 & $-0.388$ \\
\bottomrule
\end{tabular}
\end{table}

\begin{landscape}
\begin{figure}[htbp]
\centering
\includegraphics[width=0.62\linewidth, height=0.9\textheight, keepaspectratio]{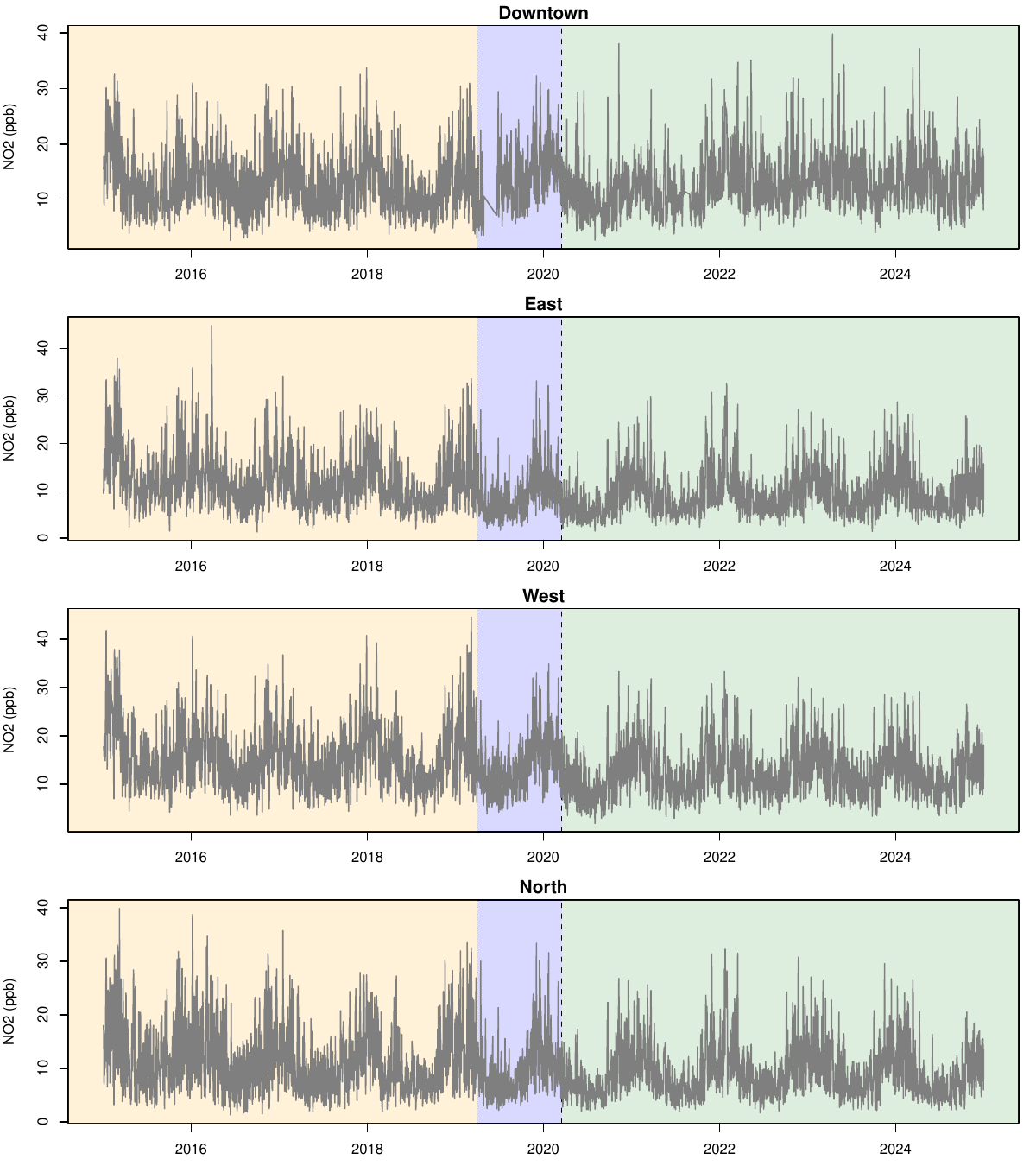}
\caption{Daily nitrogen dioxide by region, with the Drive Clean, post-removal, and
pandemic periods shaded and dashed lines at the two interruption dates.}
\label{fig:eda-no2}
\end{figure}
\end{landscape}

\begin{landscape}
\begin{figure}[htbp]
\centering
\includegraphics[width=0.62\linewidth, height=0.9\textheight, keepaspectratio]{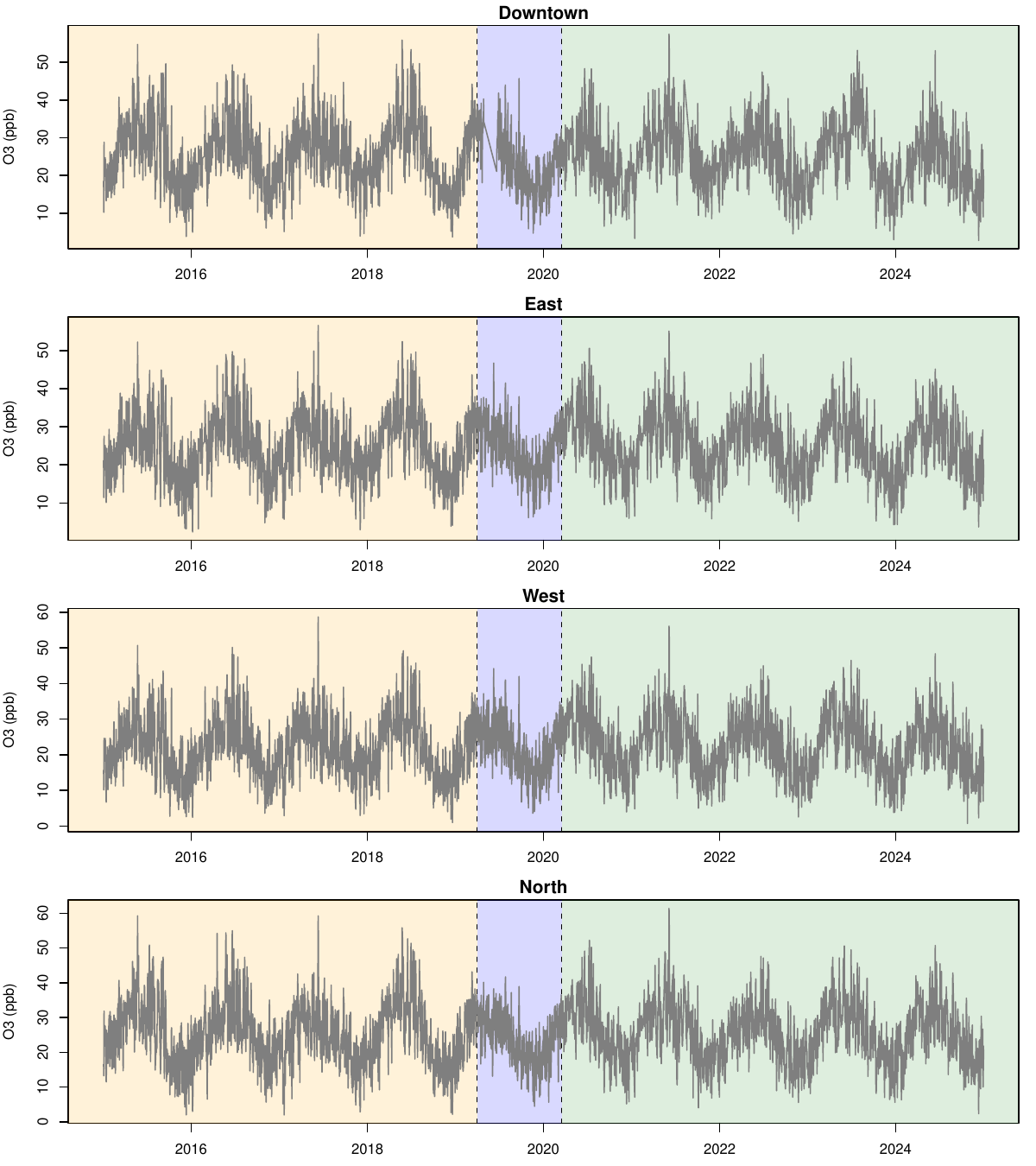}
\caption{Daily ozone by region, with the Drive Clean, post-removal, and pandemic
periods shaded and dashed lines at the two interruption dates.}
\label{fig:eda-o3}
\end{figure}
\end{landscape}

Figures~\ref{fig:eda-no2} and \ref{fig:eda-o3} show the daily nitrogen dioxide and
ozone series by region, one panel per region. In each figure three shaded periods,
with dashed vertical lines at the boundaries, mark the Drive Clean program period
before April 2019, the post-removal pre-pandemic period from April 2019 to
mid-March 2020, and the pandemic period thereafter. Nitrogen dioxide declines over
the study in the suburban regions and shows a further drop at the pandemic
boundary, while ozone shows a strong seasonal cycle and a rise during the pandemic
period.

\section{Application diagnostics}
\label{app:diag}

Table~\ref{tab:appdiag} summarizes the weighting diagnostics. The largest absolute
correlation between a confounder and NO$_2$ falls from between $0.324$ and $0.487$
before weighting to between $0.080$ and $0.149$ after single-day weighting, which
indicates that the mediator weight removes most of the measured mediator-outcome
confounding. The single-day weight retains an effective sample size between $2440$
and $2871$ of about $3500$ days, whereas the cumulative weight retains between
$765$ and $1629$, which accounts for the wider intervals and the attenuation of the
cumulative estimates. Augmented Dickey-Fuller statistics for both pollutants lie
between $-31.73$ and $-21.78$, far below the 5\% critical value near $-2.86$, so
the series are stationary and support the weak-dependence assumption used for
inference.

\begin{table}[htbp]
\centering
\caption{Application diagnostics by region: effective sample size under the
single-day (sIPW) and cumulative (Cum) weights; the largest absolute
confounder-NO$_2$ correlation before and after single-day weighting; the
concurrent-event coefficients on NO$_2$ and O$_3$ (ppb); and augmented
Dickey-Fuller statistics.}
\label{tab:appdiag}
\footnotesize
\begin{tabular}{lrrrrrrrr}
\toprule
Region & ESS sIPW & ESS Cum & max$|r|$ pre & max$|r|$ post & $C\!:$NO$_2$ & $C\!:$O$_3$ & ADF NO$_2$ & ADF O$_3$ \\
\midrule
Downtown & 2871 & 1629 & 0.324 & 0.080 & $-1.20$ & $1.82$ & $-31.73$ & $-22.00$ \\
East     & 2773 &  932 & 0.458 & 0.132 & $\phantom{-}0.06$ & $1.34$ & $-27.78$ & $-23.15$ \\
West     & 2680 &  986 & 0.473 & 0.101 & $-1.20$ & $0.06$ & $-28.76$ & $-24.15$ \\
North    & 2440 &  765 & 0.487 & 0.149 & $-0.59$ & $1.51$ & $-29.44$ & $-21.78$ \\
\bottomrule
\end{tabular}
\end{table}

%% file: references.bib
@article{baron1986moderator,
  author  = {Baron, Reuben M. and Kenny, David A.},
  title   = {The Moderator-Mediator Variable Distinction in Social Psychological Research: Conceptual, Strategic, and Statistical Considerations},
  journal = {Journal of Personality and Social Psychology},
  year    = {1986},
  volume  = {51},
  number  = {6},
  pages   = {1173--1182}
}

@inproceedings{pearl2001direct,
  author    = {Pearl, Judea},
  title     = {Direct and Indirect Effects},
  booktitle = {Proceedings of the Seventeenth Conference on Uncertainty in Artificial Intelligence},
  year      = {2001},
  pages     = {411--420},
  publisher = {Morgan Kaufmann},
  address   = {San Francisco}
}

@article{robins1992identifiability,
  author  = {Robins, James M. and Greenland, Sander},
  title   = {Identifiability and Exchangeability for Direct and Indirect Effects},
  journal = {Epidemiology},
  year    = {1992},
  volume  = {3},
  number  = {2},
  pages   = {143--155}
}

@article{imai2010general,
  author  = {Imai, Kosuke and Keele, Luke and Tingley, Dustin},
  title   = {A General Approach to Causal Mediation Analysis},
  journal = {Psychological Methods},
  year    = {2010},
  volume  = {15},
  number  = {4},
  pages   = {309--334}
}

@book{vanderweele2015explanation,
  author    = {VanderWeele, Tyler J.},
  title     = {Explanation in Causal Inference: Methods for Mediation and Interaction},
  year      = {2015},
  publisher = {Oxford University Press},
  address   = {New York}
}

@article{valeri2013mediation,
  author  = {Valeri, Linda and VanderWeele, Tyler J.},
  title   = {Mediation Analysis Allowing for Exposure-Mediator Interactions and Causal Interpretation: Theoretical Assumptions and Implementation with {SAS} and {SPSS} Macros},
  journal = {Psychological Methods},
  year    = {2013},
  volume  = {18},
  number  = {2},
  pages   = {137--150}
}

@article{vanderweele2014decomposition,
  author  = {VanderWeele, Tyler J. and Vansteelandt, Stijn and Robins, James M.},
  title   = {Effect Decomposition in the Presence of an Exposure-Induced Mediator-Outcome Confounder},
  journal = {Epidemiology},
  year    = {2014},
  volume  = {25},
  number  = {2},
  pages   = {300--306}
}

@article{tchetgen2012semiparametric,
  author  = {Tchetgen Tchetgen, Eric J. and Shpitser, Ilya},
  title   = {Semiparametric Theory for Causal Mediation Analysis: Efficiency Bounds, Multiple Robustness, and Sensitivity Analysis},
  journal = {The Annals of Statistics},
  year    = {2012},
  volume  = {40},
  number  = {3},
  pages   = {1816--1845}
}

@article{vanderweele2017timevarying,
  author  = {VanderWeele, Tyler J. and Tchetgen Tchetgen, Eric J.},
  title   = {Mediation Analysis with Time Varying Exposures and Mediators},
  journal = {Journal of the Royal Statistical Society: Series B (Statistical Methodology)},
  year    = {2017},
  volume  = {79},
  number  = {3},
  pages   = {917--938}
}

@article{lin2017gformula,
  author  = {Lin, Sheng-Hsuan and Young, Jessica G. and Logan, Roger and Tchetgen Tchetgen, Eric J. and VanderWeele, Tyler J.},
  title   = {Parametric Mediational g-Formula Approach to Mediation Analysis with Time-Varying Exposures, Mediators, and Confounders},
  journal = {Epidemiology},
  year    = {2017},
  volume  = {28},
  number  = {2},
  pages   = {266--274}
}

@techreport{zheng2012survival,
  author      = {Zheng, Wenjing and van der Laan, Mark J.},
  title       = {Causal Mediation in a Survival Setting with Time-Dependent Mediators},
  institution = {U.C. Berkeley Division of Biostatistics},
  type        = {Working Paper},
  number      = {295},
  year        = {2012}
}

@article{tchetgen2014natural,
  author  = {Tchetgen Tchetgen, Eric J. and VanderWeele, Tyler J.},
  title   = {Identification of Natural Direct Effects When a Confounder of the Mediator Is Directly Affected by Exposure},
  journal = {Epidemiology},
  year    = {2014},
  volume  = {25},
  number  = {2},
  pages   = {282--291}
}

@article{shpitser2011complete,
  author  = {Shpitser, Ilya and VanderWeele, Tyler J.},
  title   = {A Complete Graphical Criterion for the Adjustment Formula in Mediation Analysis},
  journal = {The International Journal of Biostatistics},
  year    = {2011},
  volume  = {7},
  number  = {1},
  pages   = {Article 16}
}

@book{ding2024first,
  author    = {Ding, Peng},
  title     = {A First Course in Causal Inference},
  year      = {2024},
  publisher = {Chapman and Hall/CRC},
  address   = {Boca Raton}
}

@article{robins2000marginal,
  author  = {Robins, James M. and Hern{\'a}n, Miguel {\'A}ngel and Brumback, Babette},
  title   = {Marginal Structural Models and Causal Inference in Epidemiology},
  journal = {Epidemiology},
  year    = {2000},
  volume  = {11},
  number  = {5},
  pages   = {550--560}
}

@article{lange2012simple,
  author  = {Lange, Theis and Vansteelandt, Stijn and Bekaert, Maarten},
  title   = {A Simple Unified Approach for Estimating Natural Direct and Indirect Effects},
  journal = {American Journal of Epidemiology},
  year    = {2012},
  volume  = {176},
  number  = {3},
  pages   = {190--195}
}

@inproceedings{hong2010ratio,
  author    = {Hong, Guanglei},
  title     = {Ratio of Mediator Probability Weighting for Estimating Natural Direct and Indirect Effects},
  booktitle = {Proceedings of the American Statistical Association, Biometrics Section},
  year      = {2010},
  pages     = {2401--2415},
  publisher = {American Statistical Association},
  address   = {Alexandria, VA}
}

@article{vanderweele2016practitioner,
  author  = {VanderWeele, Tyler J.},
  title   = {Mediation Analysis: A Practitioner's Guide},
  journal = {Annual Review of Public Health},
  year    = {2016},
  volume  = {37},
  pages   = {17--32}
}

@article{vanderweele2009msm,
  author  = {VanderWeele, Tyler J.},
  title   = {Marginal Structural Models for the Estimation of Direct and Indirect Effects},
  journal = {Epidemiology},
  year    = {2009},
  volume  = {20},
  number  = {1},
  pages   = {18--26}
}

@article{bernal2017its,
  author  = {Lopez Bernal, James and Cummins, Steven and Gasparrini, Antonio},
  title   = {Interrupted Time Series Regression for the Evaluation of Public Health Interventions: A Tutorial},
  journal = {International Journal of Epidemiology},
  year    = {2017},
  volume  = {46},
  number  = {1},
  pages   = {348--355}
}

@article{wagner2002segmented,
  author  = {Wagner, Anita K. and Soumerai, Stephen B. and Zhang, Fang and Ross-Degnan, Dennis},
  title   = {Segmented Regression Analysis of Interrupted Time Series Studies in Medication Use Research},
  journal = {Journal of Clinical Pharmacy and Therapeutics},
  year    = {2002},
  volume  = {27},
  number  = {4},
  pages   = {299--309}
}

@article{bhaskaran2013time,
  author  = {Bhaskaran, Krishnan and Gasparrini, Antonio and Hajat, Shakoor and Smeeth, Liam and Armstrong, Ben},
  title   = {Time Series Regression Studies in Environmental Epidemiology},
  journal = {International Journal of Epidemiology},
  year    = {2013},
  volume  = {42},
  number  = {4},
  pages   = {1187--1195}
}

@article{bojinov2019time,
  author  = {Bojinov, Iavor and Shephard, Neil},
  title   = {Time Series Experiments and Causal Estimands: Exact Randomization Tests and Trading},
  journal = {Journal of the American Statistical Association},
  year    = {2019},
  volume  = {114},
  number  = {528},
  pages   = {1665--1682}
}

@article{blackwell2018tscs,
  author  = {Blackwell, Matthew and Glynn, Adam N.},
  title   = {How to Make Causal Inferences with Time-Series Cross-Sectional Data under Selection on Observables},
  journal = {American Political Science Review},
  year    = {2018},
  volume  = {112},
  number  = {4},
  pages   = {1067--1082}
}

@book{lahiri2003resampling,
  author    = {Lahiri, S. N.},
  title     = {Resampling Methods for Dependent Data},
  year      = {2003},
  publisher = {Springer},
  address   = {New York}
}

@article{kunsch1989jackknife,
  author  = {K{\"u}nsch, Hans R.},
  title   = {The Jackknife and the Bootstrap for General Stationary Observations},
  journal = {The Annals of Statistics},
  year    = {1989},
  volume  = {17},
  number  = {3},
  pages   = {1217--1241}
}

@article{morris2019using,
  author  = {Morris, Tim P. and White, Ian R. and Crowther, Michael J.},
  title   = {Using Simulation Studies to Evaluate Statistical Methods},
  journal = {Statistics in Medicine},
  year    = {2019},
  volume  = {38},
  number  = {11},
  pages   = {2074--2102}
}

@article{dersimonian1986meta,
  author  = {DerSimonian, Rebecca and Laird, Nan},
  title   = {Meta-Analysis in Clinical Trials},
  journal = {Controlled Clinical Trials},
  year    = {1986},
  volume  = {7},
  number  = {3},
  pages   = {177--188}
}

@article{veroniki2016methods,
  author  = {Veroniki, Areti Angeliki and Jackson, Dan and Viechtbauer, Wolfgang and Bender, Ralf and Bowden, Jack and Knapp, Guido and Kuss, Oliver and Higgins, Julian P. T. and Langan, Dean and Salanti, Georgia},
  title   = {Methods to Estimate the Between-Study Variance and Its Uncertainty in Meta-Analysis},
  journal = {Research Synthesis Methods},
  year    = {2016},
  volume  = {7},
  number  = {1},
  pages   = {55--79}
}

@article{higgins2003measuring,
  author  = {Higgins, Julian P. T. and Thompson, Simon G. and Deeks, Jonathan J. and Altman, Douglas G.},
  title   = {Measuring Inconsistency in Meta-Analyses},
  journal = {British Medical Journal},
  year    = {2003},
  volume  = {327},
  number  = {7414},
  pages   = {557--560}
}

@article{sillman1999relation,
  author  = {Sillman, Sanford},
  title   = {The Relation between Ozone, {NOx} and Hydrocarbons in Urban and Polluted Rural Environments},
  journal = {Atmospheric Environment},
  year    = {1999},
  volume  = {33},
  number  = {12},
  pages   = {1821--1845}
}

@article{kleinman2000ozone,
  author  = {Kleinman, Lawrence I. and Daum, Peter H. and Imre, Dan and Lee, Yin-Nan and Nunnermacker, Linda J. and Springston, Stephen R. and Weinstein-Lloyd, Judith and Rudolph, Jochen},
  title   = {Ozone Production Rate and Hydrocarbon Reactivity in Five Urban Areas: A Cause of High Ozone Concentration in Downtown Areas},
  journal = {Geophysical Research Letters},
  year    = {2000},
  volume  = {27},
  number  = {13},
  pages   = {1867--1870}
}

@article{sicard2020amplified,
  author  = {Sicard, Pierre and De Marco, Alessandra and Agathokleous, Evgenios and Feng, Zhaozhong and Xu, Xiaobin and Paoletti, Elena and Rodriguez, Jose Jaime Diaz and Calatayud, Vicent},
  title   = {Amplified Ozone Pollution in Cities during the {COVID-19} Lockdown},
  journal = {Science of the Total Environment},
  year    = {2020},
  volume  = {735},
  pages   = {139542}
}

@article{adams2020air,
  author  = {Adams, Matthew D.},
  title   = {Air Pollution in Ontario, Canada during the {COVID-19} State of Emergency},
  journal = {Science of the Total Environment},
  year    = {2020},
  volume  = {742},
  pages   = {140516}
}

@manual{rcore2024,
  title        = {R: A Language and Environment for Statistical Computing},
  author       = {{R Core Team}},
  organization = {R Foundation for Statistical Computing},
  address      = {Vienna, Austria},
  year         = {2024}
}

@misc{driveclean2019,
  author       = {{Ontario Ministry of the Environment, Conservation and Parks}},
  title        = {Ending Drive Clean Light-Duty Vehicle Emissions Testing},
  year         = {2019},
  note         = {Light-duty passenger vehicle emissions testing ended April 1, 2019}
}
